\documentclass[manuscript]{aastex63}

\turnoffedit 

\received{12-July-2021}
\revised{04-Sep-2021}
\accepted{01-Oct-2021}
\submitjournal{ApJL}

\shorttitle{Primordial Size Distribution}
\shortauthors{OSSOS Core}

\graphicspath{{./}{figures/}}

\begin{document}

\defcitealias{2001AJ....122.1051G}{G01}
\defcitealias{2004AJ....128.1364B}{B04}
\defcitealias{2008Icar..195..827F}{Fr08}
\defcitealias{2008AJ....136...83F}{FH08}
\defcitealias{009AJ....137...72F}{FK09}
\defcitealias{2014ApJ...782..100F}{F14}

\title{OSSOS finds an Exponential Cutoff in the Size Distribution of the Cold Classical Kuiper belt}
\correspondingauthor{JJ Kavelaars}
\email{JJ.Kavelaars@nrc-cnrc.gc.ca}

\author[0000-0001-7032-5255]{J. J. Kavelaars}
\affil{Herzberg Astronomy and Astrophysics Research Centre, National Research Council of Canada, 5071 West Saanich Rd, Victoria, British Columbia V9E 2E7, Canada}
\affil{Department of Physics and Astronomy, University of Victoria, Elliott Building, 3800 Finnerty Rd, Victoria, BC V8P 5C2, Canada}
\affil{Department of Physics and Astronomy, University of British Columbia, 6224 Agricultural Road, Vancouver, BC V6T 1Z1, Canada}

\author[0000-0003-0407-2266]{Jean-Marc Petit}
\affiliation{Institut UTINAM UMR6213, CNRS, Univ. Bourgogne Franche-Comt\'e, OSU Theta F25000 Besan\c{c}on, France}

\author[0000-0002-0283-2260]{Brett Gladman}
\affiliation{Department of Physics and Astronomy, University of British Columbia, 6224 Agricultural Road, Vancouver, BC V6T 1Z1, Canada}

\author[0000-0003-3257-4490]{Michele T. Bannister}
\affiliation{School of Physical and Chemical Sciences --- Te Kura Mat\={u}, University of Canterbury,
Private Bag 4800, Christchurch 8140,
New Zealand}

\author[0000-0003-4143-8589]{Mike Alexandersen} 
\affiliation{Center for Astrophysics $|$ Harvard \& Smithsonian, 60 Garden Street, Cambridge, MA 02138, USA}

\author[0000-0001-7244-6069]{Ying-Tung Chen} 
\affiliation{Institute of Astronomy and Astrophysics, Academia Sinica; 11F of AS/NTU Astronomy-Mathematics Building, Nr. 1 Roosevelt Rd., Sec. 4, Taipei 10617, Taiwan}

\author[0000-0001-8221-8406]{Stephen D. J. Gwyn}
\affil{Herzberg Astronomy and Astrophysics Research Centre, National Research Council of Canada, 5071 West Saanich Rd, Victoria, British Columbia V9E 2E7, Canada}

\author[0000-0001-8736-236X]{Kathryn Volk}
\affil{Lunar and Planetary Laboratory, 1629 E University Blvd, Tucson, AZ 85721-0092, USA}

\begin{abstract}
The cold main classical Kuiper Belt consists of those non-resonant small solar system bodies with low orbital inclinations and orbital semi-major axes between 42.4 and 47.7 au.  These objects likely formed \textit{in situ} and the population has experienced minimal collisional modification since formation. Using the Outer Solar System Origins Survey (OSSOS) ensemble sample and characterization,  combined with constraints from deeper surveys and supported by evidence from the Minor Planet Center catalog and the Deep Ecliptic Survey,  we determine the absolute magnitude $H_r$ distribution  of the cold classical belt from $H_r\simeq5$ to 12 (roughly diameters of 400 km to 20 km). We conclude that the cold population's  $H_r$ distribution exhibits an exponential cutoff at large sizes. Exponential cutoffs at large sizes are not a natural outcome of pair-wise particle accretion but exponentially tapered power-law size distributions are a feature of numerical simulations of planetesimal formation via a streaming instability. Our observation of an exponential cutoff agrees with previous observational inferences that no large objects  ($D \gtrsim 400$ km) exist in the cold population. We note that the asymptotic slope of the $H_r$ distribution is consistent with $\alpha \sim 0.4$ and this asymptotic slope is also found in streaming instability modelling of planetesimal formation and is thus not necessarily associated with  achieving collisional equilibrium. Studies of the transneptunian region are providing the parameters that will enable future streaming-instability studies to determine the initial conditions of planetesimal formation in the  $\approx$45 au region of the Sun's protoplanetary disk.
\end{abstract}


\section{Introduction\label{sec:intro}}

Cold main classical Kuiper belt objects (KBOs) appear to be unevolved products of the initial planetesimal formation process in this region of the Solar System. 
The current number density of cold objects is such that collisions between these planetesimals are infrequent \citep[e.g.,][]{2019ApJ...872L...5G,2021AJ....161..195A}. 
The cold KBOs are known to contain a large number of loosely bound binary pairs \citep{2008ssbn.book..345N}; such pairs are very likely to be destroyed if collisions among KBOs are common \citep{2004Icar..168..409P} implying that the number density at the epoch of formation was similar to that we see today. 
The cold KBO pairs are so loosely bound that many would not have survived gravitational scattering into this zone of the solar system, implying they formed {\it in-situ} \citep{2010ApJ...722L.204P}.
Recently, observations of the cold classical KBO 486958 Arrokoth by the New Horizons mission have provided direct evidence of the low collision rate in this region \citep{2020Sci...367.6620M}.
Arrokoth impactors are dominantly cold classical KBOs and the observed low crater density, well below the crater saturation threshold, is consistent with the ancient number density of material being within a factor of a few of that currently observed in this region
\citep{2019ApJ...872L...5G,2021AJ....161..195A}.
In addition, the photometric properties of the cold belt members appear distinct from the rest of the KBOs \citep[e.g.,][]{2003ApJ...599L..49T,2017AJ....154..101P,2019ApJS..243...12S}.
The cold objects thus provide a window into the processes of planetesimal formation. 

An examination today of the $H_r$ (absolute magnitude) distribution of cold objects larger than 20~km may provide a direct measurement of the distribution that emerged from the initial planetesimal formation processes.
This population, unlike the collisionally evolved asteroid belt, never experienced the runaway growth to proto- or dwarf-planets, experienced minimal collisional erosion and has dynamical and surface properties that are distinct from the rest of the KBOs.  
The cold KBOs mass function today is the most likely to resemble the initial mass function of planetesimals. 

The outcomes of planetesimal formation modeling are reaching a point where guidance from rigorous observational constraints are needed.
The mass range in current model outputs overlaps with the well-observed range probed by the Outer Solar System Origins Survey (OSSOS).
Here we present the high-fidelity measurement of the absolute magnitude distribution of the observed cold classical Kuiper belt as determined from an ensemble of survey samples  \citep{2009AJ....137.4917K,2011AJ....142..131P,2016AJ....152..111A,2017AJ....153..236P,2018ApJS..236...18B} and associated detection characterizations (hereafter referred to as OSSOS++).

\section{The OSSOS++ cold-belt absolute magnitude distribution}

From the OSSOS++ sample \citep[see Table 3 of][for full details]{2018ApJS..236...18B} we select those 321 objects with free inclination $i_{free} < 4^\circ$ and semimajor axis in the range $42.4~\textrm{au} < a < 47.7~\textrm{au}$ to provide a relatively clean sample of the properties of cold main classical Kuiper belt objects.
\citet{2019AJ....158...49V} found the orbital parameter that best separates the cold population from the background of the main classical Kuiper belt is the inclination with respect to the $a$-dependent Laplace plane, i.e.,  $i_{free}$, and that $i_{free} < 4^\circ$ provides a reasonable split between the cold and excited populations. 
The $39.9~\textrm{au} \lesssim a \lesssim 42.4~\textrm{au}$ zone is usually included in the nominal definition of the main classical Kuiper belt.
This zone, however, was likely destabilized by the passage of the $\nu_8$ resonance during Neptune migration and KBOs on low-$i$ orbits would have been removed.
The few low-$i_{free}$ KBOs in this zone today are unlikely to have formed \textit{in situ} and for that reason we exclude them when considering the cold classicals.
Our sample criterion may exclude a small number of cold members from our analysis but ensures minimal contamination from other populations that may not have formed \textit{in situ} and would distort the view of the unevolved $H_r$ distribution.

\begin{figure}[ht]
\plotone{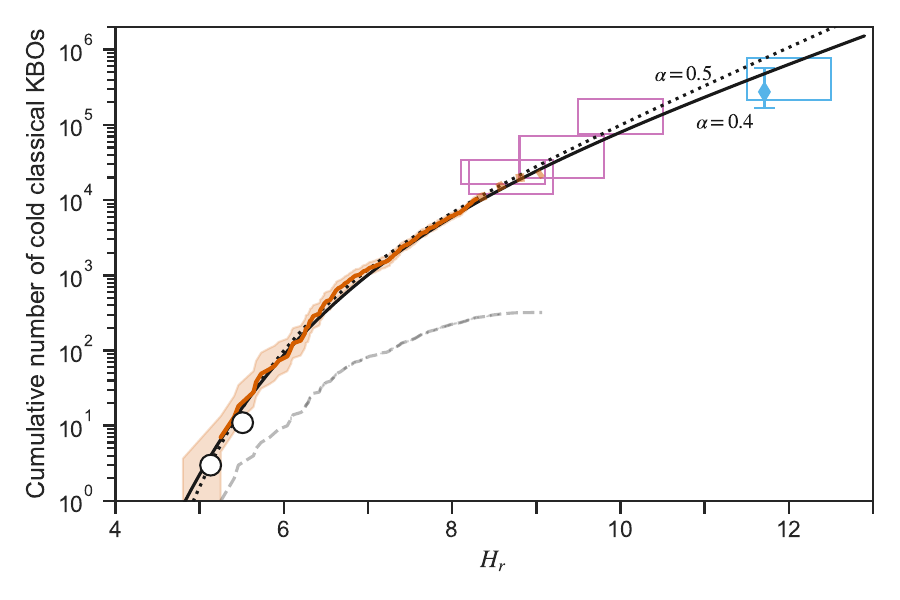}
\figcaption{$H_r$ distribution of the cold main Kuiper belt.
The grey-dash line represents the distribution of raw detections in the OSSOS++ sample.  Red-orange curve (shown as a dotted line for $H_r >8.3$ where our sample debiasing is less secure) represents the debiased OSSOS++ sample with the shading indicating the Poisson 95\% confidence range. 
The black lines represent two exponentially tapered functions matched (see Section~\ref{sec:exptap}) to the debiased OSSOS++ data, with forced large-$H_r$ (small object) asymptotic slopes (dotted: $\alpha = 0.5$; solid: $\alpha = 0.4$). 
For $H_r<9$ the two model curves are nearly identical. 
The debiased OSSOS++ measurements are well matched by the exponential taper form. 
The boxes represent literature derived estimates, see Table~\ref{tab:deep_surveys} and Section~\ref{sec:deep}.
The cyan diamond with uncertainty represents a direct debiasing of detected cold classicals in B04. 
The black open circles are located where the MPC database indicates a cumulative total of 3 ($H_r \sim 5.13$) and 11 ($H_r \sim 5.51$) main-belt cold objects.
\label{fig:Cum_H_OSSOS}}
\end{figure}

\subsection{Sample characterization}

Using the OSSOS++ characterization \citep{2018ApJS..236...18B} we debias the orbit and $H_r$ distributions of detected objects.  
The  low eccentricity of the cold KBO orbits result in these objects exploring a limited range of solar distances within a constrained phase-space volume.  
As a result of this constraint, the OSSOS++ sample of cold objects provide a complete sampling of the orbit distribution (there are no hidden or unseen populations) and we can robustly debiased the detected sample.
We consider the 4-dimensional phase space $(a, q, \sin{(i_{free})}, H_r)$ and slice it into cells small enough $(0.2~\textrm{au}, 0.2~\textrm{au}, 0.001, 0.1)$ such that the distribution of elements within a cell are likely uniform \citep[see Figure~5 of][to see the distribution of these elements]{2018ApJS..236...18B}. 
To determine the detection bias we create model objects that uniformly sample each cell's parameter range and use the OSSOS survey simulator \citep{2018FrASS...5...14L} to determine the fraction that would have been detected by the OSSOS++ surveys.
For each element cell $k$, the detection bias, $\mathcal{B}_k$, is the number of simulated orbits detected in that cell divided by the number of orbits drawn from the cell. 
For each cell we simulated the detection of 5000 objects by OSSOS++ and the value of $\mathcal{B}_k$ is 5000 divided by the number of draws from the model needed to achieve the 5000 simulated detection.
For the cold main classical belt, the values of the bias range from $\mathcal{B}_k \simeq 1/10$ for $H_r \simeq 5.5$ objects near the inner boundary of the classical region to $\mathcal{B}_k \simeq 1/100$ for the $H_r \simeq 8.3$ objects near the outer exterior of the classical region.  Although we computed bias factors for objects with larger $H_r$ values we do not use those in our analysis as the bias correction factors grow rapidly near the limit of detection.
For each of the observed cold objects in the sample, we determine the element cell, $k$, the detection belongs to and add 1/$\mathcal{B}_k$ objects to our model, with the specific elements of those model objects drawn randomly across the cell's element distribution. 
Using this procedure provides an estimate of the number of objects in the cold population required to generate the OSSOS++ detections and the distribution of the population over each of the orbit and absolute magnitude $(a, q, \sin{(i_{free})}, H_r)$ parameters.
Figure~\ref{fig:Cum_H_OSSOS} presents the resulting OSSOS++ cumulative $H_r$ distribution. 

\explain{Section added to address reviewer suggestions.}
\subsection{Other observational constraints.}

We compare with other cold KBO samples to verify our total population estimates at both the faint and bright ends of the $H_r$ distribution. 
Determination of population statistics requires carefully tracking each detected object to ensure that the orbit is accurately determined and object correctly classified. 
This tracking requires significant investment in telescope time if one is to avoid ephemeris bias entering the sample \citep{2006Icar..185..508J,2008ssbn.book...59K}.
This also places significant constraint on the faintness of objects that are allowed into a particular survey as the cost of tracking will rapidly increase.
The OSSOS++ sample provides a high-quality sampling of the cold Kuiper belt due to the near 100\% effectiveness in tracking detections to obtain high-quality orbits.  
The desire to achieve this high success rate in tracking, however, also limited the flux range accessible to the survey at both the bright and faint ends of the $H_r$ distribution.
Other surveys which do not have precise orbit and distance estimates for all their objects, however, can be used to estimate the cold classical $H$- magnitude distribution and extend the flux range explored.  
We find that these additional samples agree well with the results measured via the OSSOS++ sample, providing a verification of the absolute calibration of our study.

\subsubsection{Deep studies\label{sec:deep}}
\explain{Section title changed to remove the word Luminosity}

\citet{2001AJ....122.1051G,2004AJ....128.1364B,2008Icar..195..827F,2008AJ....136...83F,2009AJ....137...72F}
(hereafter G01, B04, Fr08, FH08, and FK09) performed deep `pencil-beam' surveys to detect faint TNOs and published their detection efficiency functions. 
The imprecise determination of heliocentric distance at detection in G01, Fr08, FH08 and FK09, however, prevents direct conversion from observed brightness to $H_r$ (absolute) magnitudes. 
We carefully examined the sample of detections from each of these projects.
The VLT part of G01 did not yield any detections and we keep only the CFHT component of that project (G01/CFHT). 
The inclinations in the Blanco part of Fr08 are insufficiently constrained to allow selection between cold and excited objects in that sample, thus we utilized only the CFHT component from that project (Fr08/CFHT).
The details provided in FH08 and FK09 are sufficient to allow use of their full sample of detected cold KBOs.
For each survey, we examined the published detection efficiency curves and determined the limiting apparent magnitude $M_r$ (all surveys used here reported limits in $r$) up to which the efficiency of detection, $\eta$, is roughly constant (see Table~\ref{tab:deep_surveys}). 
For each survey we:
\begin{itemize}
    \item count the number $n$ of cold objects (estimated\footnote{We cannot compute $i_{free}$ as the semimajor axis of the orbits of the objects are, generally, unknown.} inclination $\le 4^\circ$) brighter than $M_r$
    \item estimate the actual number of objects present in the field of view of the survey and brighter than $M_r$ as $n/\eta$
    \item determine the fraction of the full cold population that is in the field of view of a survey at any given time, $F$, using the CFEPS orbit model \citep{2011AJ....142..131P}.
    \item compute an estimate of the implied full population brighter than $M_r$: $N(m<M_r) = n/(F \eta)$.
\end{itemize}
The 95\% confidence range for $N$ is computed from the 95\% Poisson confidence range for $n$ and listed as $N_{-}$ for the lower end of the confidence range and $N_{+}$ for the upper in Table~\ref{tab:deep_surveys}.
To add these density estimates to our $H_r$ distribution we must convert the observed apparent magnitude limit ($M_r$) to an absolute ($H_r$) value, which requires knowledge of the distance to the sources.  
As the precise distances of the individual detections are not known we determine a plausible range of $H_r$ values by adopting the 95\% range of distances of the cold objects from the CFEPS model \citep{2011AJ....142..131P,2014ApJ...782..100F} (40.3 to 51~au).
Using this approach we determine $H_r(51)$  and $H_r(40.3)$  (using $H_{51} = M_r - 17.03$ and $H_{40.3} = M_r - 16.00$) as representative of the range $H_r$ values that each survey was sensitive to at the detection limit ($M_r$). 

B04 provides precise distance estimates and even rough orbital elements and the characterization of the detection efficiency. 
Thus, for B04, we also debiased the 3 detections of that survey following the same procedure as for OSSOS++. 
For B04, the full debiasing results in a population estimate at the largest $H_r$ of the detected B04 objects that is compatible with the estimate using the process outlined in the preceding paragraph (Figure~\ref{fig:Cum_H_OSSOS}).  

We discuss implications for the observed $H_r$ distribution in Sections~\ref{sec:exptap}.

\begin{deluxetable}{lrrrrrrrr}
\tablecaption{Deep Surveys.  
\label{tab:deep_surveys} 
}

\tablehead{
\colhead{Survey\tablenotemark{a}} & \colhead{$\eta$} & \colhead{$M_r$} & \colhead{$n_{cold}$} &  \colhead{1/$F$} & \colhead{$H_{51}$} & \colhead{$H_{40.3}$} & \colhead{$N_-$} & \colhead{$N_+$}
}
\startdata
\citet{2008AJ....136...83F}       & 0.88 & 25.1 & 30 &    930 &  8.1 &  9.1  &  22300 &   45300 \\
\citet{2008Icar..195..827F} CFHT  & 0.97 & 25.2 & 11 &   1726 &  8.2 &  9.2 &   11000 &   35000 \\
\citet{2001AJ....122.1051G} CFHT  & 1.00 & 25.8 &  3 &  11700 &  8.8 &  9.8 &   10733 &   94100 \\
\citet{2009AJ....137...72F}       & 0.95 & 26.5 & 10 &  11594 &  9.5 & 10.5 &   67000 &  224500 \\
\citet{2004AJ....128.1364B}    & 1.00 & 28.5 &  3 & 115325 & 11.5 & 12.5 &  125700 & 1011100 \\
\enddata
\tablenotetext{a}{If a telescope is listed, the sample is restricted to that particular portion of the study.}
\end{deluxetable}

\subsubsection{Inventory of the Brightest Cold Classicals}

The now nearly complete, and sparse, inventory of the lowest H (largest) cold classical KBOs provides a further constraint on the size distribution.
We select from the Minor Planet Center (MPC) database KBOs consistent with the $a$ and $i_{free}$ cuts given above and with good orbits (2 oppositions or more, 6 observations or more, and MPC orbit uncertainty parameter less than or equal to 6).
The MPC database provides the visual absolute magnitude $H$ which we convert into $H_r$ using $\left<H - H_r\right> = 0.19$, the mean value for the cold objects from OSSOS++ present in the MPC database.
We then determine the number of known cold classical KBOs with $H_r < 5.1$ (3) and $H_r < 5.5$ (11)
and include this as an estimate of the cumulative $H_r$ distribution; see Figure~\ref{fig:Cum_H_OSSOS}.
These MPC-derived population estimates are nearly identical to the OSSOS++ based estimate of the total numbers at these $H_r$ values, confirming the global population estimates obtained by our debiasing.
That the OSSOS++ estimate of the total population of cold objects is the same size as the currently known sample, suggests that, as previously noted \citep[e.g.][]{2011AJ....142...98S}, the MPC database has reached (near) completeness for $H$ around 5-6 mag.
The expectation that the brightest members of the cold population would be in this range was also noted in Figure~7 of \citet{2004AJ....128.1364B} and is apparent in Figure~9 of \citet{2014ApJ...782..100F}.
The expectation of completeness in the MPC sample is also coherent with the reported detection of large cold objects by Pan-STARRS, which surveys the whole ecliptic. 
During the period 2010-2014, Pan-STARRS found 7 of the 20 largest objects in the cold belt region defined above. 
Since 2014 no new large cold objects have been reported by Pan-STARRS, despite continuous operation.
An absence of $H_r < 4$ cold objects was also predicted a decade ago based on the CFEPS project \citep[see Sec.~5.1.1 of][]{2011AJ....142..131P} whose sample is included in OSSOS++.
The total inventory of known cold population objects with $H_r<5.5$ matches precisely the prediction from the OSSOS++ $H_r$-distribution, appears to be complete, and is small in number.

\section{An exponentially tapered $H_r$ distribution.}
\label{sec:exptap}

The shape of the cold population $H_r$ distribution presented in Figure~\ref{fig:Cum_H_OSSOS} is inconsistent with two-component power-law\footnote{We sometime use the term power-law to refer to the exponential forms such as $N(<H) \propto 10^{\alpha H}$ as the underlying mass distribution is, in-fact, a power-law form and $H$ is used as a proxy for that quantity, and referring to exponentially tapered exponential functions becomes exponentially confusing, but does reflect the exponential complexity of reality.} fits. 
Using our de-biased model we imposed various two-component power-laws \citep[see][]{2014ApJ...782..100F} onto our model and compared the resulting orbit and $H_r$ distributions  with our observed sample.  
No acceptable models \citep[rejected by AD test statistic at more 99\% confidence, see][ for details of our statistical procedure]{2018FrASS...5...14L} were found.  
Moving to a multi-component power-law could provide a solution, but is difficult to physically motivate. 
The shape is also not consistent with the rolling power-law utilized in \cite{2004AJ....128.1364B}.  
The steep and continuously evolving shape at the bright-end of the $H_r$ distribution is inconsistent with these representations. 

A tapered power-law form appears to be emerging as a preferred functional form for planetesimal mass distributions.
Disk instability mechanisms, and the streaming instability (SI) process in particular, have recently become highly favored solutions to overcoming various physical barriers in planetesimal formation \citep[e.g.,][]{1972epcf.book.....S,2007Natur.448.1022J,2019PASP..131g2001L} and enable planetesimal formation to proceed more rapidly at lower surface densities, like those in the primordial cold Kuiper belt.  
Independent groups have investigated the initial mass function of planetesimals resulting from the SI \citep{2015SciA....1E0109J,2016ApJ...822...55S,2017A&A...597A..69S,2019ApJ...883..192A,2019ApJ...885...69L,2020A&A...638A..88L,2021MNRAS.500..520R}. 
Due to limitation in the simulated mass resolution, the low-mass end of the  mass function that emerges from SI models is not strongly constrained.
The existing works, however, exhibit similar-shaped mass distributions that a power-law can roughly approximate at the small mass end but require a rather sharp, exponential, cut-off at the large mass end. 
Several functional forms to fit the mass distributions have been proposed.
We select the exponentially tapered (\citet{2017A&A...597A..69S}  or variably tapered, \citet{2019ApJ...885...69L}) power-law form.
According to Bayes criterion these forms provide better matches to the SI outcomes than simpler, single parameter functions \citep{2017A&A...597A..69S,2019ApJ...885...69L}.

The exponentially tapered power-law mass distribution \citep{2017A&A...597A..69S} can be transformed into an $H_r$ distribution assuming a constant albedo and density, and spherical shape:
\begin{equation}\label{eqn:LF}
    N(<H_r)=  10^{\frac{3}{5}\alpha_{SI}(H_r-H_o)} \; \exp\left[ 10^{-\frac{3}{5}\beta_{SI} (H_r-H_B)}\right]
\end{equation}
\explain{equation form re-expressed to be more intuitively understandable.}
where $N$ will be the total population number
and $H_o$ is a normalization, $\alpha_{SI}$ is the asymptotic slope at large $H_r$, $\beta_{SI}$ is the strength of the exponential tapering and $H_B$ is the $H_r$ value at which the exponential taper begins to dominate.
The $\alpha$ in the single exponential $H_r$ distribution, $N(<H_r) = 10^{\alpha(H_r-H_o)}$, is given by $\alpha = \frac{3}{5}\alpha_{SI}$ and can be seen as related to the faint/small object exponent when considering multi-component exponential distributions. 

The OSSOS++ sample confirms the general shape of an exponential taper and can be used to determine the strength of that tapering ($\beta_{SI}$).
The OSSOS++ sample is, however, limited to the large $H_r$ end where our debiasing factors are small; the value of $\alpha_{SI}$ is not robustly constrained by the OSSOS++ sample. 
Crater counts on Pluto and Charon \citep{2019Sci...363..955S}, the observed sizes of Jupiter-family comets \citep{2012Icar..218..571S} and results from deep surveys shown in Figure~\ref{fig:Cum_H_OSSOS} indicate that $\alpha \simeq$0.3--0.5 faint-ward of $H_r \sim 9$. 
In Figure~\ref{fig:Cum_H_OSSOS} we present fits of Equation~\ref{eqn:LF} with fixed values of $\alpha \in \{0.4, 0.5\}$ ($\alpha_{SI} \in \{0.66, 0.83\}$), with the value of $\beta_{SI}$ and the other free parameters determined using maximum likelihood MCMC parameter exploration via the $emcee$ package \citep{2013PASP..125..306F} over the range $H_r \in \{5.0,8.3\}$
The OSSOS++ estimates of $\beta_{SI} \in \{0.42^{+0.12}_{-0.16}, 0.59^{+0.13}_{-0.27}\}$, 
$H_o \in \{-2.6^{+0.4}_{-0.9},0.0^{+0.2}_{-0.4}\}$
and $H_B \in \{8.1^{+1.7}_{-0.6},7.1^{+0.9}_{-0.4}\}$
provide a remarkably smooth match to our debiased observations and our estimates of the value of $\beta_{SI}$ are steeper than, but consistent with, the range of values found in SI modeling \citep[0.28--0.37; e.g.][]{2017A&A...597A..69S}.

\subsection{Comparison with published distributions.}
\explain{Added section to compare with literature results.}

\begin{figure}
    \centering
    \plotone{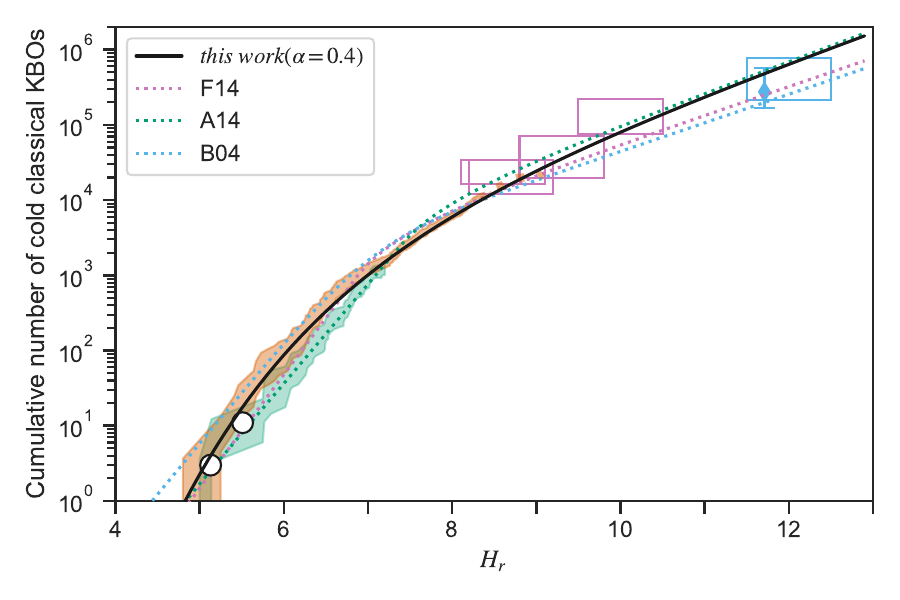}
    \caption{As in Figure~\ref{fig:Cum_H_OSSOS} the red-orange region represents the debiased OSSOS++ sample.  The green shaded area represents the debiased detections from the Deep Ecliptic Survey (DES) \citep[][]{2014AJ....148...55A} with the green dashed line their best-fit double exponential. Also shown are the best fits from \citet[][cyan dotted line]{2004AJ....128.1364B} and \citet[magenta dashed line]{2014ApJ...782..100F}.  The curves have been scaled to reflect difference in survey filters
    and for differences in selection function for cold classical KBOs. In particular we use $(r-R) = 0.25$ \citep{2006A&A...460..339J} for 
    $(V-R)=0.6$ 
    cold classical KBOs and we scale the apparent magnitude distribution given in B04 using a fixed distance of 42~au to transform from $r$ to $H_r$.  The A14 total population is slightly low compared to OSSOS++ sample, this may be due to tracking losses reported in A14.  The F14 fit has been scaled to  match the OSSOS++ sample at $H_r=8$ as we were uncertain of the scaling from the surface density reported in F14 and the absolute total numbers reported here.}
    \label{fig:comparison}
\end{figure}

Figure~\ref{fig:comparison} presents our measured $debiased$ $H_r$ distribution along with a number of results from the literature. 
The double or broken exponential forms presented in Figure~\ref{fig:comparison} do not provide as compelling a match to the observations as the tapered exponential.  
There is good agreement that asymptotic small object exponential must be around $\alpha \sim 0.4$ but the forms do not provide a good match to the data at small $H_r$ which exhibits a continuously steepening slope. 
The OSSOS++ data is consistent with the bright end exponential slopes over some limited range of $H_r$ then drops away from those curves for smaller $H_r$ and larger $H_r$.
The bright/large object end of the $H_r$ distribution is not single sloped value and thus double exponential functions are not a good match.

\section{Discussion\label{sec:discuss}}

The $H_r$ distribution of the OSSOS++ sample clearly demonstrates an exponentially tapered shape.
 OSSOS  was designed to be an absolutely calibrated survey that could be debiased to measure intrinsic absolute distributions inside each dynamical group, with the group membership based on high-precision orbits.
These orbits permit the computation of free inclinations and the exclusion of resonant objects enabling the selection of relatively uncontaminated sample of cold belt members.
We find the debiased $H_r$ distribution of this sample is inconsistent with a single power-law at the bright end (see
Figures~\ref{fig:Cum_H_OSSOS} and \ref{fig:comparison}).
The functional form of the cold component $H_r$ distribution is well represented by an exponential taper of the type seen in numerical simulations of streaming instability driven planetesimal formation.

The OSSOS++ derived $H_r$ distribution matches well onto independent constraints at both ends of the distribution.
At the faint (large $H_r$) end the debiased OSSOS++ sample connects smoothly to the faintest pencil-beam studies.
Although each individual study is difficult to map to a precise $N(<H_r)$ value, the ensemble of deep apparent-magnitude studies, collectively, match OSSOS++, implying that the asymptotic form of a single exponential from $H_r > 9 $ down to at least $H_r \approx 12$ does not violate known constraints.
The cratering records on Pluto and Charon \citep{2017Icar..287..187R,2019Sci...363..955S} and Arrokoth \citep{2020Sci...367.3999S}, however, indicate at least one further transition to an even shallower exponent beyond $H_r > 17$.
The OSSOS++ sample provides strong constraints in the $H_r\simeq$5--8.3 range; for $H_r>8.3$, where our detected sample drops off, a constant slope in log-space appears plausible for several more magnitudes.
We highlight that within the \textbf{hot} component there is evidence for a knee or divot \citep{2013ApJ...764L...2S,2016AJ....151...31S,2016AJ....152..111A} feature near $H_r$=8.5 with a similar shallow slope for $H_r > 9$.
On the large object end, the cumulative distribution of the  intrinsically brightest known cold-classical KBOs fall directly on the OSSOS++ curve and its bright-ward extrapolation.
This close match 
confirms the very steep nature of the $H$-magnitude distribution of the largest objects, and implies that the inventory of these largest objects is essentially complete.
These connections to independent constraints at $H_r$=5 and 9  (with no tuning) gives confidence that in OSSOS++ we have an absolutely calibrated survey and that the $H$-magnitude distribution's shape between these two ends  is correctly represented by our debiased measurement.

The existence of the exponential taper also resolves some literature confusion regarding the measured exponent of the KBO $H_r$ distribution.
Much of the historical literature fit a single exponential to the apparent magnitude $N(<M) \propto 10^{\alpha M}$. 
Because of the finite sky area and the small number of objects detected in any given survey, even fitting a simple exponential was challenging due to a lack of dynamic range.
Generally, larger-area surveys were shallower, while deeper surveys made up for their smaller area via the steep $H_r$ distribution to end up with comparable (but small) samples.
Magnitude distribution shape estimates done in apparent magnitude space (because they lack the precise distance estimates required to translate apparent to absolute magnitude)  `blur' the $H$-magnitude distribution, resulting in slope estimates that are a function of the depth of the survey; a similar conclusion was reached in \citet[][]{2014ApJ...782..100F}.

Papers that attempted to combine surveys to enlarge the apparent magnitude and dynamic range of observational constraints tended to average this out to an intermediate slope \citep[e.g.][]{1998AJ....116.2042G,2008Icar..195..827F}.
With an even larger apparent magnitude range it became clear that a single exponential could not represent the data, and double/rolling  exponent forms were introduced.
Magnitude distribution studies near the solar system's invariable plane (which are thus dominated by low-inclination cold-component objects) exhibited a change of $\alpha$ to shallower values fainter than $m_r\simeq$24--25 (thus $H_r\sim$8-9 in the main belt), when using a rolling \citep{2004AJ....128.1364B}, double \citep{2009ApJ...696...91F} or broken power-law \citep{2009AJ....137...72F} size distributions.

Although computationally convenient, there is no physical motivation for a broken power-law being the correct functional form.  
Numerical simulations of the streaming instability, however, appear to naturally produce the exponentially tapered form \citep{2017A&A...597A..69S,2019ApJ...885...69L}.
The OSSOS++ sample demonstrates that this functional form is an excellent representation of the cold-component $H$-magnitude distribution.
Much of the discussion of what value of $\alpha$ best matches the actual $H_r$ distribution and where to put a break to attempt to mimic the slope evolution, appears to be due to trying to model an exponential taper by combining multiple exponential distributions.

Having such examples of the danger of over-imposing a functional form on reality, we note that although this tapered exponential is clearly impressively similar to the distribution derived from the OSSOS++ sample, numerical simulations \citep[e.g.][]{2019ApJ...885...69L} show that there can be smaller features  superposed on this dominant form which, in differential space, manifest themselves as local `knees' (broken or double exponents)  or `divots' \citep{2013ApJ...764L...2S}.
In particular, the weak relative under-abundance just past $H_r\simeq7$ present in previous data sets \citep[e.g.][]{2014AJ....148...55A, 2014ApJ...782..100F} may be real, in addition to a proposed knee near $H_r\simeq8.4$
in the dynamically hot populations \citep[e.g.][]{2016AJ....152..111A,2018AJ....155..197L}.

The SI modeling prediction of the shape of the $H$-magnitude distribution at small sizes (large $H$) is not yet firmly established.
The value of the asymptotic power-law slope seen at small sizes in simulations \citep{2017ApJ...847L..12S,2016ApJ...822...55S,2019ApJ...883..192A,2021MNRAS.500..520R} may result from resolution effects. 
\citet{2019ApJ...885...69L} demonstrate that as resolution is improved, what initially appeared to be a roll-over to a `single-$\alpha$' asymptotic form, continues to evolve and the transition to the asymptotic form appears to occur at ever smaller sizes.
Interestingly, the existing estimates of these asymptotic limits \citep[see Figure~10 in][for example]{2019ApJ...883..192A} are not very far from the often-suggested value of $\alpha \simeq 0.4 \pm 0.1$  for $H_r \gg 8$  \citep{2004AJ....128.1364B,2008AJ....136...83F,2010Icar..210..944F,2014ApJ...782..100F}.
This allows the possibility that this value of $\alpha$ is then set by the \textbf{formation} size distribution and the similarity of the observed $H_r$ distribution slope to the collisional equilibrium value is not evidence of collisional equilibrium having been achieved.
The small object size distribution of the cold classical Kuiper Belt appears to be unaltered over the age of the Solar System and to preserve a shape consistent with SI planetesimal formation down to of order kilometer scale.

The comparison with current modeling of SI driven planetesimal formation is not, however, without a significant hurdle.  
Although the shape of the distribution is compelling, the inferred mass ranges are not a good match.
If the current cold component is indeed a relatively unevolved population, then the current surface density may be taken as a proxy of the density at formation. 
Based on our estimate of the total cold population (Figure~\ref{fig:Cum_H_OSSOS}) we can infer the surface density at the time of the SI process.  
To estimate this density we convert our $H_r$ distribution to a mass distribution by assuming a constant albedo of 0.15 and object bulk density of 500 kg/m$^3$ \citep{2020Sci...367.3999S} and then spread the inferred total mass into a 2~au wide ring centered at 43~au.  
The resulting estimated primordial surface density in solids is $\Sigma_p \sim 5\times10^{-5}\; \textrm{g}/\textrm{cm}^2$.
This is 100 times lower than the mass scale required for most SI modeling to produce $D \sim 400~\textrm{km}$ objects \citep[e.g.,][]{2019ApJ...883..192A}.
The number densities inferred from the current population are far too low to be consistent with SI processes forming the sizes of KBOs in our observed size distribution, which we claim follows the shape seen in SI modeling!  
Conversely the number density at the time of planetesimal formation implied by SI modeling results appears to be significantly higher than that inferred from the currently observed population. 

There appear to be at least the following possible issues to consider: 
\begin{itemize}
\item The current density could be many factors lower than at the time of planetesimal formation. 
This appears unlikely as any process that removed significant mass  would very likely have disrupted the binary KBO population we see today.  
The cratering record seen on Arrokoth is fully consistent with the low total populations reported here \citep[see][]{2019ApJ...872L...5G,2019AGUFM.P33I3535S,2021AJ....161..195A} indicating that any period of high number density would have been very rapidly removed which would have implications for the orbit distribution in the cold belt
\citep{doi:10.1146/annurev-astro-120920-010005}.
Thus, it appears unlikely that the surface density at $\sim43~\textrm{au}$ was significantly higher than today. 

\item The SI process is not directly responsible for the production of the objects we see today, but they formed instead after the SI process via particle-particle interactions. 
A 2~au wide ring would contain many hundreds to thousands of SI cells \citep[e.g.][]{2019ApJ...885...69L}, and perhaps the planetesimals from these cells coalesce to form the largest bodies.
Here, again, the currently observed density makes this appear improbable as the inferred particle-particle encounter rate would be very low (thus the low numbers of craters) making the rate of planetesimal growth so slow that the largest sized objects would have not yet formed.  
In addition, there is no reason that the observed $H_r$ distribution of the post-growth populations would then so closely resemble that which emerges from the SI processes.
\item The models of SI are incomplete and the process happens on scales and at densities that have not yet been fully modelled.  
This appears unlikely to be the case as the modeling is done in scale free units and then the mass scales are set by imposing a density.  
\item The classical Kuiper belt was much more tightly confined, radially and azimuthally, at the time of SI driven planetesimal formation and then rapidly dispersed shortly after the planetesimals emerged.  
This \textit{ad hoc} solution has some appeal as many planet forming disks exhibit density enhancements \citep[e.g.][]{2013Sci...340.1199V} and the shearing out of particles that might form in such dense regions would be quite rapid compared to collisional time-scales.  
To achieve the required density enhancement, however, would require a concentration that is 100 times that seen today. Are such density enhancements feasible? 
\end{itemize}

Some caution is thus warranted because what is observed today in the cold-classical Kuiper belt is, in this paradigm, the end state of streaming instability plus later accretion and erosion.
If the latter two processes are indeed negligible (or could be successfully modeled) then the shape of the cold classical belt $H$-magnitude distribution becomes a {\it direct} measure of the outcome of the planetesimal formation process from the streaming instability.
Future numerical work can then in principle constrain the protoplanetary disk's parameters  (surface density, viscosity, etc.) as one targets reproducing the observed distribution in the preserved cold classical belt.

Regardless of a possible link with the SI process, our analysis of the OSSOS++ sample has provided a robust high-fidelity measure of the $H_r$ distribution of the cold component of the Kuiper belt.  This analysis is enabled by the precise characterization of the OSSOS++ surveys.  The derived shape exhibits an exponentially tapered form and, from the current evidence, is representative of the initial distribution resulting from planetesimal formation.

\section*{acknowledgments}
JJK and BG acknowledge support of the Natural Sciences and Engineering
Research Council of Canada.  This work was supported by the Programme
National de Plan\'etologie (PNP) of CNRS-INSU co-funded by CNES.  This
research made use of the Canadian Advanced Network for Astronomy
Research (CANFAR) and the facilities of the Canadian Astronomy Data
Centre operated by the National Research Council of Canada with the
support of the Canadian Space Agency.  KV acknowledges support from
NASA (grants NNX15AH59G and 80NSSC19K0785) and NSF (grant
AST-1824869).  Based on observations obtained with MegaPrime/MegaCam,
a joint project of CFHT and CEA/DAPNIA, at the Canada-France-Hawaii
Telescope (CFHT) which is operated by the National Research Council
(NRC) of Canada, the Institut National des Sciences de l'Univers of
the Centre National de la Recherche Scientifique (CNRS) of France, and
the University of Hawaii. The observations at the CFHT were performed
with care and respect from the summit of Maunakea which is a
significant cultural and historic site.

\bibliography{bibliography}{}
\bibliographystyle{aasjournal}

\listofchanges 
\end{document}